\documentstyle[12pt,psfig]{article}
\def\reference{\parskip 0pt\par\noindent\hangindent 0.5 truecm}

\begin{document}
%
%
\title{Rotation of Early B-type Stars in the Large Magellanic Cloud \\
- The role of evolution and metallicity.}
%


\author{Stefan C. Keller $^{1}$
} 

\date{}
\maketitle

{\center
$^1$ Research School of Astronomy and Astrophysics, Private Bag, Weston, ACT 2611, Australia\\ stefan@mso.anu.edu.au\\[3mm]
}

%
\begin{abstract}
I present measurements of the projected rotational velocities of a sample of
100 early B-type main-sequence stars in the Large Magellanic Cloud. This is
the first extragalactic study of the distribution of stellar rotational
velocities. The sample is drawn from two sources: a sample derived from the
vicinity of the main-sequence turnoff of young clusters (ages
1-3$\times10^{7}$yrs), and a sample from the general field. I find the cluster
population exhibits significantly more rapid rotation than that seen in the
field. I have drawn analogous Galactic cluster and field samples from the
literature. Comparison of these samples reveals the same effect. I propose the
observed difference between cluster and field populations can be explained by
a scenario of evolutionary enhancement of the surface angular momentum over
the main-sequence lifetime. A comparison is made between the cluster and field
populations of the LMC and the Galaxy in order to explore the effects of
metallicity. This shows that the LMC stars are more rapid rotators than their
Galactic counterparts.

\end{abstract}


\bigskip

\section{Introduction}
This paper aims to establish the distribution of projected rotational
velocities ($v$sin$i$) for a sample of early B-type stars in the Large
Magellanic Cloud (LMC). I explore the effects of the parameters age and
metallicity on the observed distribution of rotational velocities. The
motivation for this study arises from recent stellar evolutionary models that
incorporate stellar rotation. This modeling shows that the rotational history
is vital to the interpretation of the evolutionary phase and elemental
abundances seen in massive stars. Heger, Langer \& Woosley (2000) and Meynet
\& Maeder (2000) show that axial rotation can bring about in a natural way the
degree of extension to the convective core in massive stars as discerned from
cluster H-R diagrams (e.g.\ Keller et al.\ 2000, 2001a) and the
mass-luminosity relation of Cepheids (Keller \& Wood 2002).

Models incorporating rotation also predict mixing of internally processed
material to the surface during main-sequence (MS) evolution (the faster the
initial rotation the greater the degree of mixing). Standard evolution models
(e.g. Bressan et al. 1993), on the other hand, do not. In the Magellanic
Clouds (MCs), nitrogen has been shown to be over-abundant in many supergiants
of spectral types B to K (e.g. Venn 1999; Lennon et al.\ 1996) and numerous MS
B stars (Rolleston et al.\ 1996; Korn et al.\ 2000; Korn et al.\ 2002) which
is indicative of mixing of CNO cycled material. The dispersion in N abundance
is significantly larger than that seen in the Galaxy (Mc Erlean et al.\ 1999;
Venn 1995; Gies \& Lambert 1992). This has been tentatively attributed to the
presence of a greater number of rapid rotators within the lower metallicity
environs of the MCs.

At present, direct observational studies of the dependence of stellar rotation
on metallicity are limited. Within the Galaxy the work of Burki \& Maeder
(1977) examined the variation of mean $v$sin$i$ of early B stars with
galactocentric distance, but the range of the latter was only several kpc
which, given the galactic metallicity gradient, is insufficient to draw firm
conclusions regarding the role of metallicity. Recent work by Royer et al.\
(2003) has revisited this work extending it to a broader sample of Galactic
clusters with inconclusive results.

Indirect observational evidence for more rapid rotation of massive stars in
metal-poor sites like the MCs is derived from examination of the population of
Be stars. Be stars show Balmer emission arising from a circumstellar disk of
material surrounding a rapidly rotating stellar photosphere. They form a
population with which to trace stellar rotation (assuming that the onset of
the Be phenomenon itself is not governed by metallicity but rather the
underlying distribution of rotational velocities). Maeder et al.\ (1999)
examine the fraction of Be stars (i.e.\ N(Be)/N(B)+N(Be) ) against metallicity
for a sample of clusters with relevant ages (1-3$\times$10$^{7}$ years,
corresponding to the maximum occurrence of Be stars). Maeder et al. find a
strong anti-correlation between the Be fraction and metallicity. This leads
the authors to conclude that rotation rates are higher amongst metal-poor
stars.

The role of age in determining the resultant distribution of $v$sin$i$ was
first suggested by Abt \& Hunter (1962) in their study of the brighter members
of the Trapezium and Pleiades clusters. This was elaborated upon by Wolff et
al.\ (1982) and Guthrie (1984) in their studies which contrasted the
distribution of rotational velocities seen in young clusters, older
associations and field stars. The distribution of rotational velocities in the
older associations and the field share a high proportion of very slow rotators
compared with the young cluster sample.

Keller et al.\ (1999, 2000, 2001b) have used the Be star population to explore
the age dependence of the distribution of rotational velocities. We have
observed that the Be fraction amongst the young cluster population rises
towards the luminosity of the MS terminus, whereas in the field it retains a
relatively uniform distribution in luminosity. This difference between cluster
and field populations can, we argue (Keller et al.\ 2001b), be brought about
from an evolutionary enhancement in the rotational velocity over the later
portion of the MS lifetime.

It is the aim of this paper to present direct measurements of the projected
rotational velocities for a sample of early B-type MS stars within young
cluster and field environments of the LMC, together with analogous Galactic
data from the literature, to investigate directly the dependence of $v$sin$i$
on both age and metallicity. Section 2 presents our observations and method
for the determination of projected rotational velocities. Section 3.1 compares
the distribution of rotational velocities in the cluster and field
environments of the Galaxy and the LMC. Section 3.2 compares the Galactic and
LMC cluster populations to explore metallicity effects. Finally in Section 4
I present a discussion of the physical causation of the observed dependence
of the distribution of projected rotational velocities on age and metallicity.

\section{Observations and data reduction}
\label{se:obs}

I have defined two sample sets: (1.) A cluster sample -- drawn from the
population in the vicinity of three young populous clusters in the LMC,
namely, NGC 1818, NGC 2004 and NGC 2100. Our targets were selected from the
WFPC2-based photometry of Keller et al.\ (2000). On the basis of
this photometry I have selected those stars on the MS with
$14.0<V<16.0$. (2.) A field sample -- from the vicinity of NGC 2004 and
1818. These regions were selected to form the basis of our sample because of
their richness and apparently small differential reddening (Keller, Wood \&
Bessell 1998).

Medium resolution (0.6\AA/px) spectra of the sample were obtained using the
Double-Beam Spectrograph (DBS) on the SSO 2.3m telescope between 4-6 March
1999, 25-16 November 1999 and 10-12 January 2001. The spectra consist of two
simultaneously recorded non-overlapping segments: blue (3800$-$4800\AA) and
red (6200$-$6800\AA). The observations made use of aperture plates machined to
locate the target stars. In general the density of the targets precluded the
use of slits for each object, rather holes of appropriate size for typical
seeing at the site (2") were made. Sky correction was latter made through the
use of slits at the extremities of the field. Wavelength calibration was made
through CuAr spectra interleaved with the observations.

\subsection{Determination of rotational velocities}

To determine the projected rotational velocity, $v$sin$i$, I have used a
$\chi^2$ minimisation technique which locates an optimal match between the
observed spectrum and rotationally broadened synthetic spectra. The grid of
synthetic profiles of stellar H Balmer and HeI lines by Gonzales Delgado \&
Leitherer (1999) is used. This grid of models spans 50000$\ge$
T$_{\rm{eff}} \ge$4000 K and 0.0$\ge \rm{log} g \ge$ 5.0. For
T$_{\rm{eff}} > 25000$ K NLTE models are used to compute the synthetic
spectra, for cooler stars, Kurucz LTE models are used.

To create the rotationally broadened spectra the synthetic spectra were first
convolved with the instrumental response function and then with a broadening
function which is based upon that of Gray (1976): model stellar surface is
divided into 40 segments and the emergent doppler-shifted intensity from each
is calculated. This is then integrated to yield the flux at each wavelength.

This approach assumes that the resultant line profile is independent of
viewing angle. Physically, the surface gravity (and hence temperature) of a
rotating star varies with latitude as a result of gravity darkening (von
Zeipel 1924). As a consequence the contribution of the rapidly rotating cooler
equatorial regions to the observed spectrum is reduced. Townsend et al.\
(2004) present a survey of the behaviour of line width with vsini including
viewing angle dependent gravity darkening. For instance, a star rotating at
95\% of the critical velocity (i.e.\ the velocity at which the equatorial
escape velocity drops to zero) seen pole on using the method we apply here
leads to an underestimation of $v_{e}sini$ by 20\%. The underestimation of
$v_{e}sini$ is a strong function of $v_e$ such that by $v_e$=300kms$^{-1}$ for
a B2 star this has fallen to a few percent. The neglect of gravity darkening
underlies all rotational velocity standards and previous studies of rotational
velocity distributions. For this reason the present work maintains this
simplification.

\begin{figure}
\begin{center}
\psfig{file=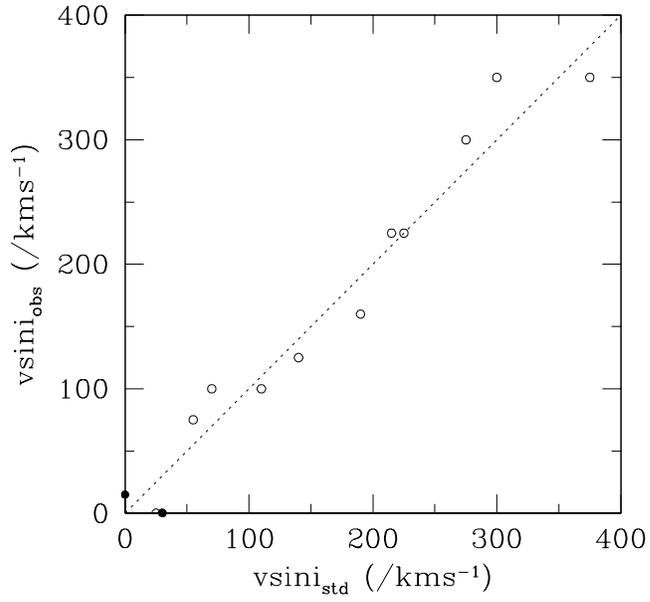,height=8cm}
\caption[]{Comparison between projected rotational
velocities of Slettebak et al. (1975, open circles) and those of the present
study. Solid points show the positions of B15 and B30 in NGC 2004 from Korn et
al.\ 2000.}
\label{compvsini}
\end{center}
\end{figure}

\begin{figure}
\begin{center}
\psfig{file=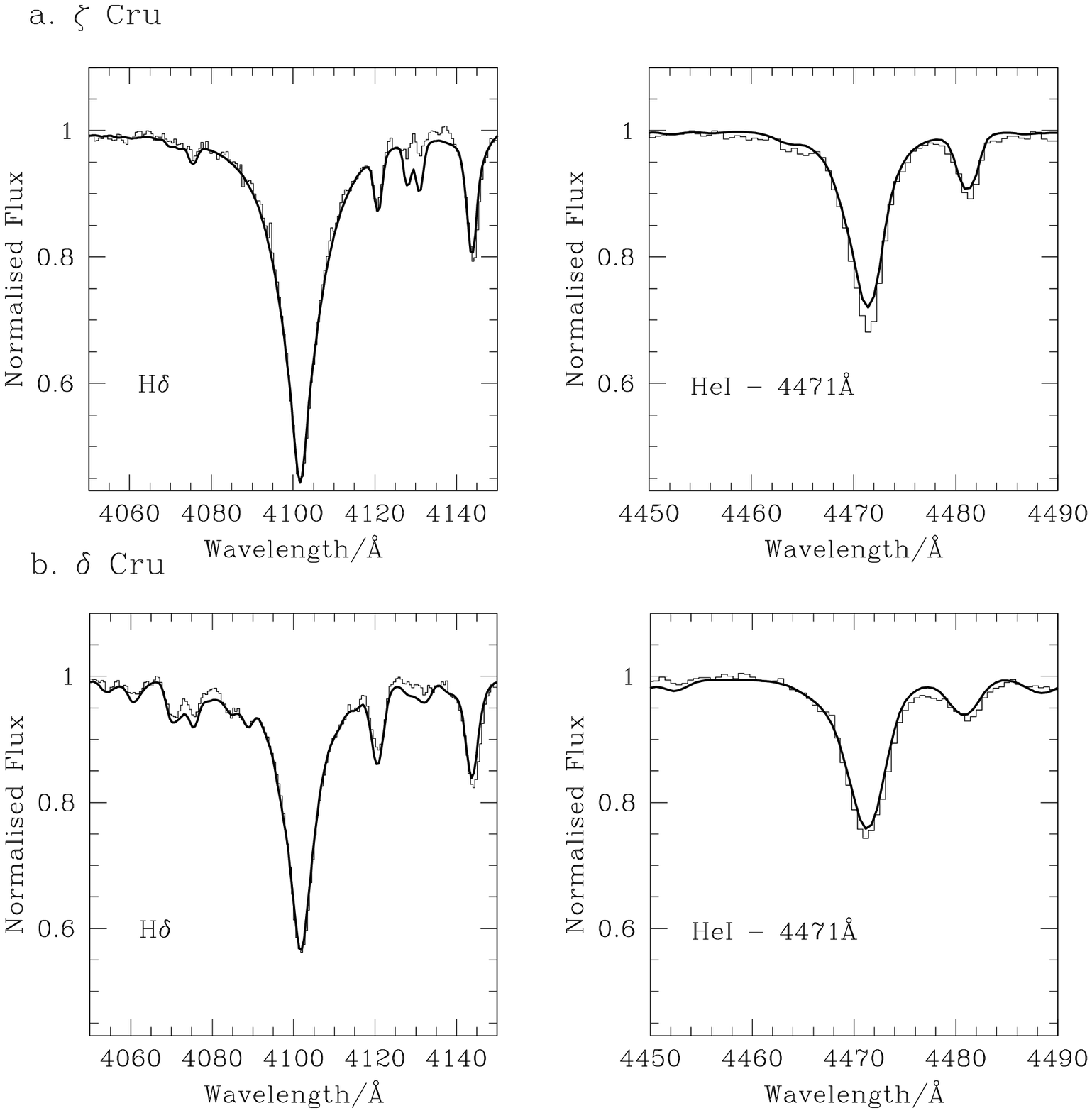,height=8cm}
\caption[]{Examples of the fits between artificially broadened theoretical
profiles (thick line) and observed stellar spectra for rotational velocity
standards; {\bf{a.}} $\zeta$ Cru (B2.5V, $v$sin$i$=70kms$^{-1}$) and
{\bf{b.}} $\delta$ Cru (B2IV, $v$sin$i$=140kms$^{-1}$)} 
\label{stdfitfig}
\end{center}
\end{figure}

I have used a wavelength range of 3990\AA$\,$ to 4550\AA$\,$ for the purposes
of finding an optimal match. The spectra is fit as one contiguous piece. An
optimal solution is sought in $v$sin$i$, T$_{\rm{eff}}$, logg parameter
space. The spectra of a series of rotational velocity standards (Slettebak et
al. 1975) of early to mid B spectral type were obtained. Figure
\ref{compvsini} shows the comparison between the $v$sin$i$ determined by
means of our parameter fit procedure and that of Slettebak et
al.~(ibid). Figure \ref{stdfitfig} shows a typical fit to the rotational
velocity standards. Figure \ref{compvsini} does not show any systematic
difference between these results and those of Slettebak et al.

\begin{figure}
\psfig{file=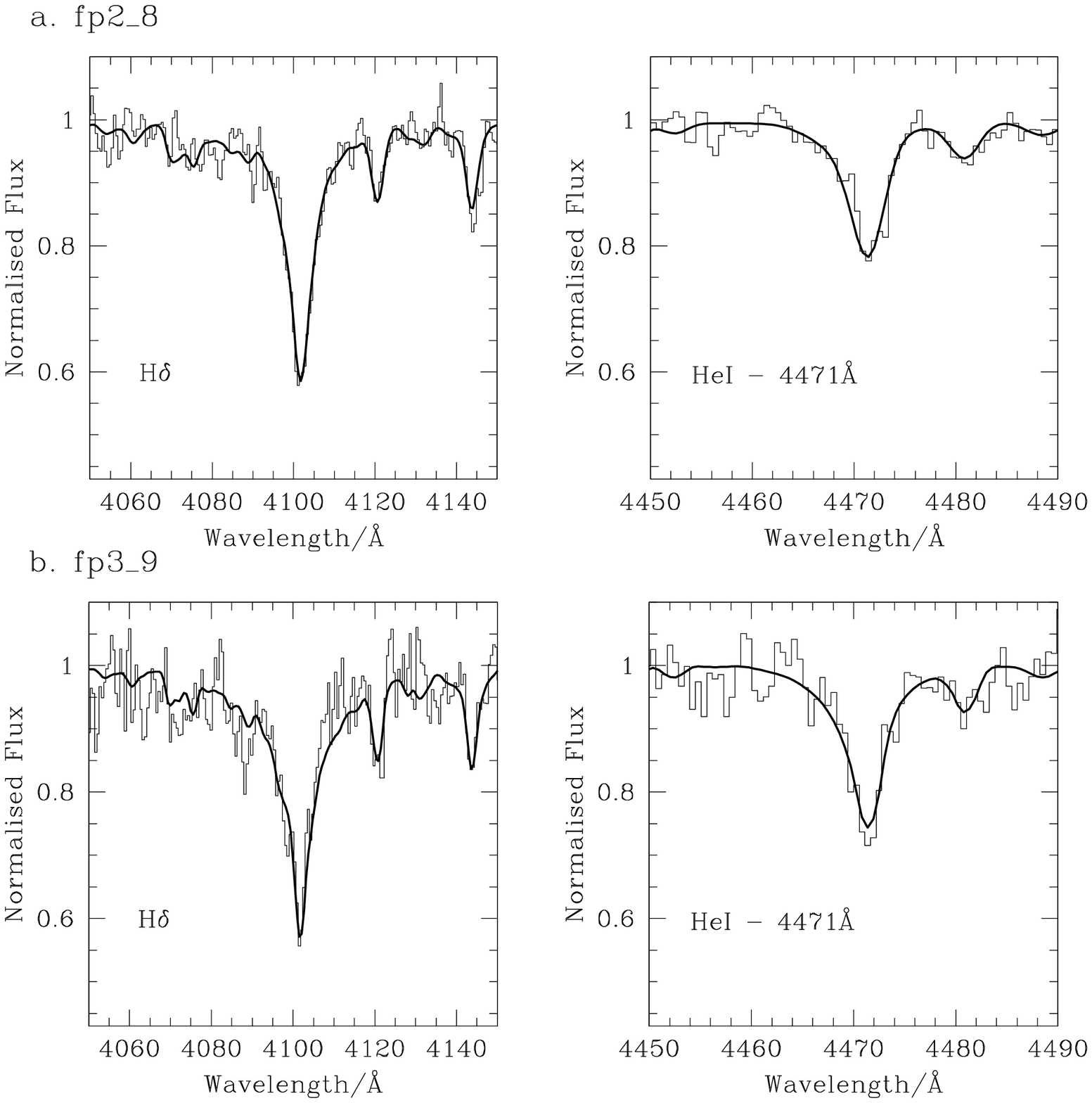,height=8cm}
\caption[]{Examples of rotationally broadened line profile fits for LMC field
stars. Plot {\bf{a.}} shows fp2\_8 ($V$=13.83 and $v$sin$i$=190kms$^{-1}$) and
{\bf{b.}} shows fp3\_9 ($V$=15.89 and $v$sin$i$=50kms$^{-1}$).}
\label{fitfig}
\end{figure}

The signal-to-noise ratio (S/N) of the spectra of our LMC sample is low, a
typical 3hr exposure results in a S/N=30 on a $V$=15.5 star. On the basis of
simulations (discussed below) I set a lower limit to the S/N of our
sample of 20 ($V \sim 16$). Table \ref{table1} \& \ref{table2} reports the
$v$sin$i$ for our sample. Some typical fits are shown in figure \ref{fitfig}.

The uncertainties in our estimated $v$sin$i$ are a function of both S/N and
$v$sin$i$. I have carried out a series of trials using model spectra which
are rotationally broadened and then degraded to a given S/N. I have then
examined the $v$sin$i$ returned by our routine from repeated trials. Slow
rotators are perturbed by a low S/N. A S/N of 20 introduces an uncertainty of
$\pm$50 kms$^{-1}$ at a $v$sin$i$ of 0 kms$^{-1}$ and systematically returns a
higher $v$sin$i$. This overestimation of $v$sin$i$ at low S/N results
primarily from the radial velocity correction which must be made for each star
during fitting. At low S/N the line-centers of the significant H and He lines
are less distinct and an optimal fit may occur with either a slightly higher
or lower radial velocity than the underlying systemic radial velocity. A S/N
degraded line seen off-center appears as a broader line and this results in
the systematic shift of the slowest rotators to higher $v$sin$i$.

\begin{figure}
\psfig{file=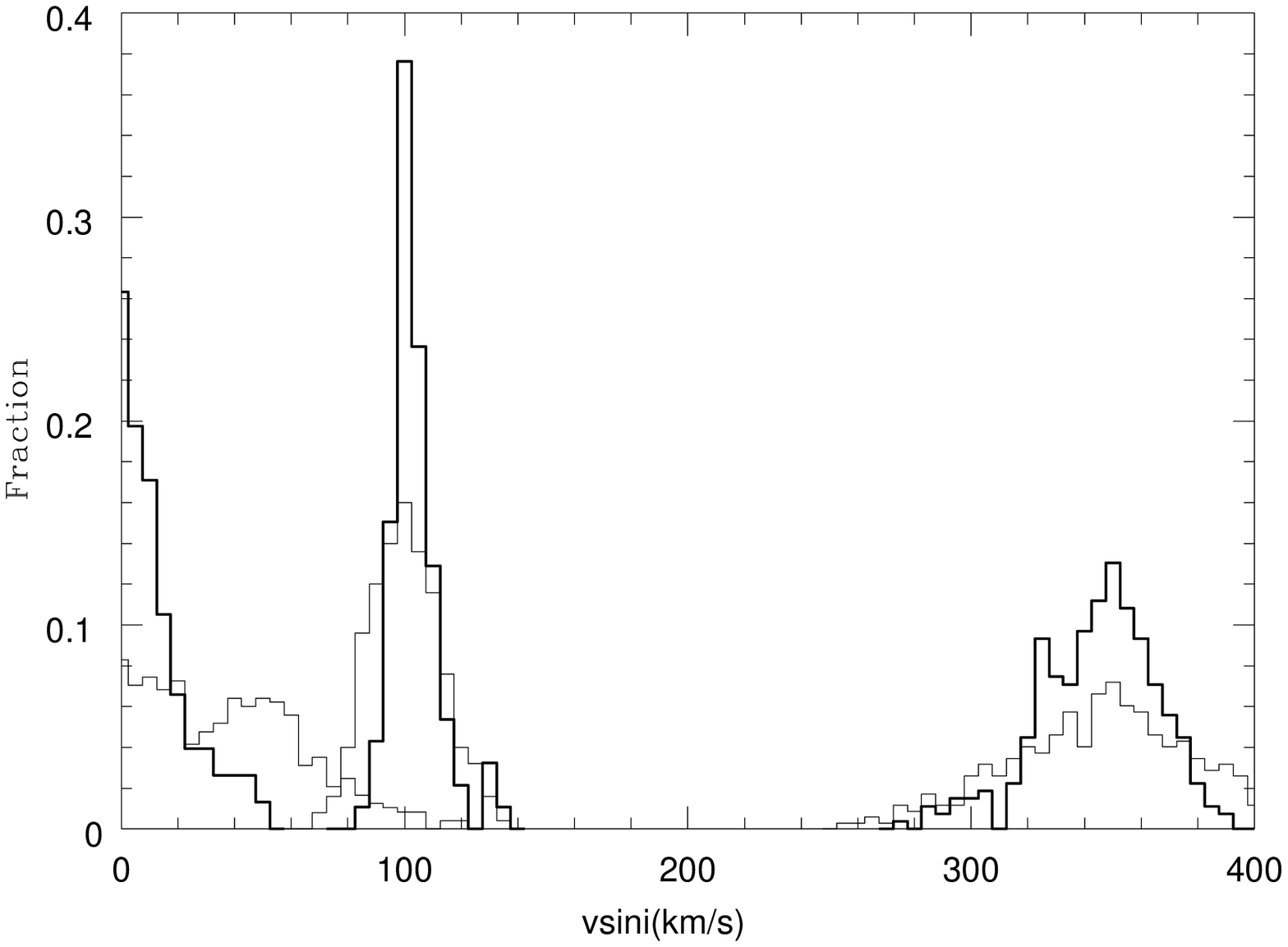,height=8cm}
\caption[]{Histograms of the $v$sin$i$ returned from our fitting routine from
three models with different input $v$sin$i$ (0, 100, 350 kms$^{-1}$ resp.) and S/N. The
thick line shows the distribution recovered from a S/N=50 and the thinner line
from S/N=20.}
\label{snfig}
\end{figure}

As $v$sin$i$ is increased the importance of this effect diminishes. At
$v$sin$i$=100 kms$^{-1}$ and S/N=20 the uncertainty in $v$sin$i$ falls to
$\pm$25 kms$^{-1}$ with no significant systematic shift. The uncertainty grows
again to $\pm$75 kms$^{-1}$ at $v$sin$i$=350 kms$^{-1}$ (S/N=20) as the line
depth is reduced but again with no significant systematic shift. This
behaviour is summarised in figure \ref{snfig}. By describing in detail the
systematic behaviour of our technique at low S/N it is possible to
statistically remove it as described in section 3.2.

\begin{table}
\caption[]{$v$sin$i$ for the sample of main-sequence stars drawn from the
LMC clusters NGC 1818, 2004 and 2100. $V$ magnitudes are taken from Robertson
(1974). $v$sin$i$ with additional colon indicate those objects for
which determination of $v$sin$i$ have considerable uncertainty ($\pm$75
kms$^{-1}$).}
\begin{center}
\begin{tabular}{lcclcc}
\hline
Star& $V$ & $v$sin$i$&Star& $V$ & $v$sin$i$\\
& & (kms$^{-1}$)& &  & (kms$^{-1}$)\\
\hline
NGC2004:C10 & 15.90 & 190 & NGC1818:B34 & 14.27 & 125 \\
NGC2004:C9  & 15.76 & 150 & NGC1818:A31 & 14.89 & 0 \\
NGC2004:C8  & 14.88 & 130 & NGC1818:C24 & 16.21 & 38: \\
NGC2004:B15 & 14.18 & 15  & NGC1818:B39 & 15.89 & 260 \\
NGC2004:B12 & 15.31 & 187 & NGC1818:C25 & 15.74 & 0: \\
NGC2004:B18 & 14.77 & 190 & NGC1818:C29 & 15.59 & 0 \\
NGC2004:B38 & 14.06 & 300 & NGC1818:C27 & 15.97 & 260 \\
NGC2004:D10 & 15.11 & 150 & NGC1818:D12 & 13.74 & 0 \\
NGC2004:D8  & 15.37 & 263 & NGC1818:C11 & 16.10 & 300: \\
NGC2004:D3  & 15.77 & 130 & NGC1818:B24 & 15.05 & 260 \\
NGC2004:D15 & 15.10 & 105 & NGC1818:D6 & 15.57 & 180 \\
NGC2004:D13 & 13.36 & 150 & NGC1818:D4 & 15.83 & 300: \\
NGC2004:D16 & 13.66 & 225 & NGC2100:D30 & 14.29 & 260 \\
NGC2004:B9  & 13.48 & 0   & NGC2100:D28 & 14.91 & 120 \\
NGC2004:B24 & 14.68 & 130 & NGC2100:D31 & 14.89 & 225 \\
NGC2004:C13 & 14.90 & 265: & NGC2100:D26 & 15.29 & 187 \\
NGC2004:B30 & 13.83 & 15  & NGC2100:C9 & 13.78 & 15 \\
NGC2004:B28 & 15.25 & 300 & NGC2100:B45 & 14.25 & 0 \\
NGC2004:C16 & 14.69 & 105 & NGC2100:B20 & 13.71 & 120\\
NGC2004:D7  & 15.04 & 280 & NGC2100:C30 & 14.93 & 90 \\
NGC2004:D1  & 12.02 & 15  & NGC2100:C13 & 13.85 & 105 \\
NGC1818:C10 & 16.12 & 188 & NGC2100:C15 & 13.68 & 30 \\
NGC1818:D15 & 16.21 & 150: & NGC2100:C27 & 15.09 & 225 \\
NGC1818:B28 & 16.13 & 225: & NGC2100:C20 & 13.56 & 0 \\
NGC1818:B29 & 15.67 & 225 & & & \\       
\hline
\end{tabular}
\end{center}
\label{table1}
\end{table}

\begin{table*}
\caption[]{$v$sin$i$ for the sample of main-sequence B stars drawn from the
LMC field. $V$ photometry is that of the authors.}
\begin{center}
\small
\begin{tabular}{lcccclcccc}
\hline
Star& RA & Dec & $V$ &$v$sin$i$&Star& RA & Dec & $V$ &$v$sin$i$\\
& (J2000) & (J2000) &  &{\footnotesize{kms$^{-1}$}}&& (J2000) & (J2000) &  &{\footnotesize{kms$^{-1}$}}\\
\hline
fp1\_1 & 05:34:10.07 & -67:03:12.5 & 14.99 & 225 & fp2\_13 & 05:32:53.80 & -66:57:20.1 & 14.62  & 75\\
fp1\_2 & 05:33:48.80 & -67:02:03.2 & 15.97 & 180: & fp2\_14 & 05:32:44.15 & -66:56:36.3 & 14.83  & 75\\
fp1\_3 & 05:34:03.63 & -67:01:50.5 & 14.69 & 100 & fp3\_1 & 05:28:24.39 & -68:15:21.8 & 15.53 & 90\\
fp1\_4 & 05:33:55.94 & -67:04:29.2 & 15.74 & 0 & fp3\_2 & 05:29:04.88 & -68:13:46.5 & 14.02 & 130\\
fp1\_5 & 05:34:18.87 & -67:03:20.0 & 15.73 & 175 & fp3\_3 & 05:29:36.93 & -68:02:52.7 & 15.19 & 240\\
fp1\_6 & 05:34:01.92 & -67:02:41.2 & 15.92 & 0: & fp3\_4 & 05:29:43.25 & -68:19:04.2 & 15.42 & 150\\
fp1\_7 & 05:34:06.28 & -67:02:35.5 & 15.12 & 100 & fp3\_5 & 05:29:57.43 & -68:10:38.3 & 15.90 & 15\\
fp1\_8 & 05:33:53.33 & -67:02:00.2 & 15.92 & 175 & fp3\_6 & 05:30:12.32 & -68:18:32.0 & 14.09 & 50\\
fp1\_9 & 05:33:46.18 & -67:01:30.0 & 15.62 & 260 & fp3\_7 & 05:30:40.69 & -68:02:52.5 & 15.88 & 75\\
fp1\_10 & 05:33:39.99 & -67:01:14.4 & 15.85 & 340 & fp3\_8 & 05:30:44.98 & -68:06:23.9 & 15.89 & 20\\
fp1\_11 & 05:34:01.48 & -67:00:40.0 & 14.86  & 110 & fp3\_9 & 05:30:58.87 & -68:15:39.5 & 15.86 & 50\\
fp1\_12 & 05:33:59.25 & -67:00:33.1 & 14.91  & 125 & fp3\_10 & 05:31:00.15 & -68:09:03.2 & 15.33 & 20\\
fp1\_13 & 05:34:05.85 & -67:00:18.5 & 15.98  & 190 & fp3\_11 & 05:31:32.10 & -68:18:56.6 & 15.99 & 250\\
fp1\_14 & 05:34:00.17 & -66:59:57.0 & 15.77 & 225 & fp3\_12 & 05:31:37.05 & -68:10:59.5 & 15.99 & 0\\
fp2\_1 & 05:32:57.32 & -67:00:35.8 & 15.89 & 190 & fp3\_13 & 05:31:49.94 & -68:08:43.7 & 15.46 & 260\\
fp2\_2 & 05:33:09.98 & -67:00:27.7 & 15.71 & 260 & fp4\_1 & 05:28:26.30 & -68:12:26.9 & 15.44 & 0\\
fp2\_3 & 05:33:16.88 & -67:00:20.2 & 15.06 & 0 & fp4\_2 & 05:28:37.01 & -68:10:41.4 & 15.93 & 180\\
fp2\_4 & 05:33:04.07 & -66:59:35.1 & 13.99  & 0 & fp4\_3 & 05:29:41.34 & -68:22:16.2 & 15.96 & 0\\
fp2\_5 & 05:32:41.76 & -66:59:17.8 & 15.63  & 225 & fp4\_4 & 05:30:18.30 & -68:19:08.3 & 15.61 & 0\\
fp2\_6 & 05:32:50.91 & -66:59:13.8 & 15.71 & 0 & fp4\_5 & 05:30:40.49 & -68:00:30.6 & 15.77 & 20\\
fp2\_7 & 05:32:33.72 & -66:59:06.2 & 14.91 & 110 & fp4\_6 & 05:31:18.35 & -68:10:50.7 & 15.18 & 0\\
fp2\_8 & 05:32:53.60 & -66:59:02.2 & 13.82  & 190 & fp4\_7 & 05:31:32.87 & -68:02:23.8 & 15.03 & 80\\
fp2\_9 & 05:32:40.64 & -66:58:25.5 & 15.98  & 0: & fp4\_8 & 05:31:48.81 & -68:10:33.0 & 15.10 & 300\\
fp2\_10 & 05:32:43.23 & -66:58:13.6 & 15.93  & 110: & fp4\_9 & 05:32:10.98 & -68:08:15.9 & 15.92 & 60\\
fp2\_11 & 05:33:16.46 & -66:57:58.1 & 15.69  & 260 & fp4\_10 & 05:32:13.38 & -68:15:30.7 & 15.36 & 0\\
fp2\_12 & 05:32:46.78 & -66:57:42.6 & 15.01  & 0 & & & & & \\
\hline
\end{tabular}
\end{center}
\label{table2}
\normalsize
\end{table*}

\section{Distribution of rotational velocities}
\label{se:vsini}

\subsection{Comparison of cluster and field populations}

\begin{figure}
\psfig{file=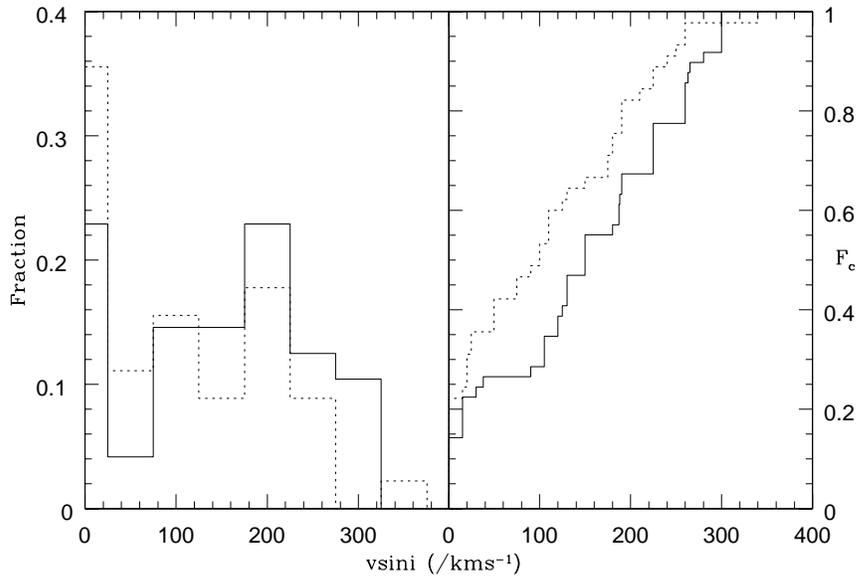,height=8cm}
\caption[]{The distribution of rotational velocities from the 49 LMC cluster
stars (solid line) and 51 field stars (dotted) shown on the left as a
histogram and to the right as the non-parametric cumulative distribution.}
\label{vsinidist}
\end{figure}

Figure \ref{vsinidist} compares the distribution of rotational velocities
from LMC cluster and field samples. The non-parametric two-sample
Kolmogorov-Smirnov (K-S) test is conducted to test the validity of the
proposition that both samples are derived from the same parent
distribution. Such a test reveals that the cluster and field are unlikely to
arise from a similar parent distribution at 2$\sigma$ significance
(Prob.(D$>$D$^{+}$)=4\%; D$^{+}$=0.25, N$_{1}$=49, N$_{2}$=51).

To extend our analysis to a discussion of the role of metallicity in the
distribution of rotational velocity I have constructed an analogous sample
from the Galactic field. Our LMC sample has a limit of $V\sim$16 imposed by
the attainable S/N. Due to the high inclination angle and the essentially
disk-like morphology of the LMC we can regard the sample as effectively
volume-limited. Assuming a distance modulus to the LMC of 18.45 our sample
extends to a ${\rm{M}}_{V}$=-3. This corresponds to a spectral type of B2 on
the MS (Zorec and Briot 1991).

The Galactic field sample is drawn from the Bright Star Catalogue (Hoffeit \&
Jaschele 1982:BSC) and the Supplement to the Bright Star Catalogue (Hoffeit et
al.\ 1988:SBSC). I have drawn from the catalogue stars of spectral types
B0-B2 and of luminosity class III-V. The BSC+SBSC sample is magnitude-limited
at $V$=7.1. In order to compare with the volume-limited sample from the LMC we
first must reduce the apparent number of stars within each spectral type and
luminosity class division to a volume-limited distribution. This must be done
otherwise the most luminous stars (i.e.\ early spectral type and high
luminosity class) will receive a disproportionate weighting (Mamquist bias).

I shall consider a common volume defined by the B0III stars. The
sample-filling factor $\epsilon$ for each spectral type and luminosity class
is given in Zorec and Briot (1997, their table 4) for the BSC+SBSC sample. The
corrected histogram distribution of $v$sin$i$, $h_{cor}(v$sin$i)$ is
constructed from the sum of the histograms for the individual spectral types
and luminosity classes, $C$, scaled by $\epsilon$:

\begin{equation}
h_{cor}(v{\rm{sin}}i) = \sum_{i} \epsilon_{i} \cdot h(C_{i}:v{\rm{sin}}i)
\end{equation}

where $\epsilon$ ranges from 4.5 for B2V to 1.0 for B0III. The net result is a
distribution which is strongly weighted by the input of the intrinsically most
numerous members, that is, those of B2V.

A distribution of projected rotational velocities was drawn from a sample of
young Galactic clusters (see table \ref{rot:galsample}). The chosen clusters
are rich clusters with analogous ages to those of the LMC cluster
sample. Stars within 2 magnitudes of the MS turnoff (B0-2V-III) were selected
for the distribution.

\begin{table}
\caption[Sample of young galactic clusters.]{Input to the Galactic cluster $v\,$sin$\:i$ distribution.}
\begin{center}
\begin{tabular}{lccl}
\hline
Cluster & log age & stars w. $v\,$sin$\:i$ & Reference\\
\hline
h+$\chi$ Per. & 7.0 & 30 & Slettebak (1968)\\
NGC3766 & 7.4 & 10 & Bernacca \& Perinotto (1971)\\
NGC2439 & 7.1 &  17 & Bernacca \& Perinotto (1971)\\
IC 4665 & 7.5 &  18 & Mermilliod (2000)\\
Sco. OB2 & 7.1 & 46 & Brown \& Verschueren (1997)\\
NGC 663 &  7.2 & 30 & Bernacca \& Perinotto (1971)\\
\hline
\end{tabular}
\end{center}
\label{rot:galsample}
\end{table}

\begin{figure}
\psfig{file=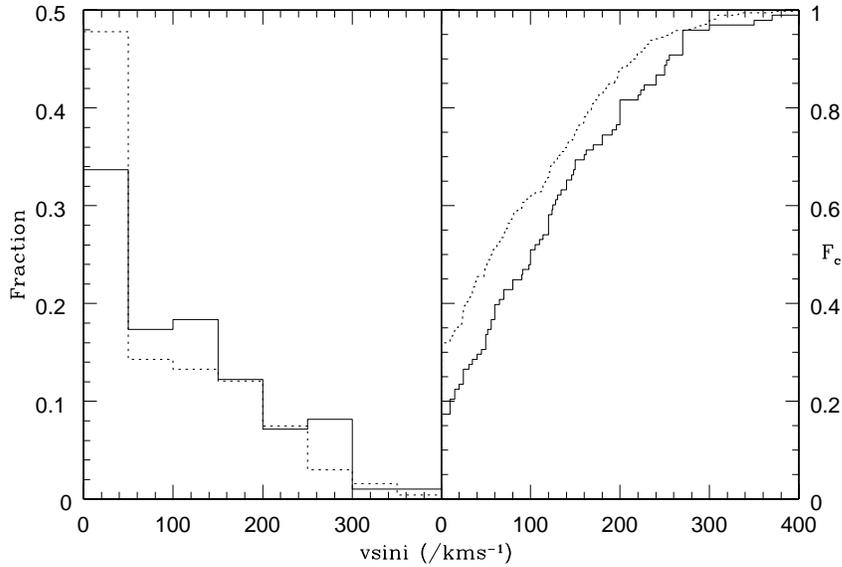,height=8cm}
\caption[]{The distribution of rotational velocities from the Galactic cluster
(solid line) and field (dotted) shown on the left as a histogram and to the
right as the non-parametric cumulative distribution.}
\label{kstestgcf}
\end{figure}

The distributions of rotational velocities for the two samples are compared in
figure \ref{kstestgcf}. The results of a K-S test on the above distributions
shows that the Galactic cluster stars are significantly faster rotators than
their field counterparts (Prob.(D$>$D$^{+}$)=2\%, D$^{+}$=0.165, N$_{1}$=310,
N$_{2}$=151).

\subsection{Comparison of Galactic and LMC populations}

\begin{figure}
\psfig{file=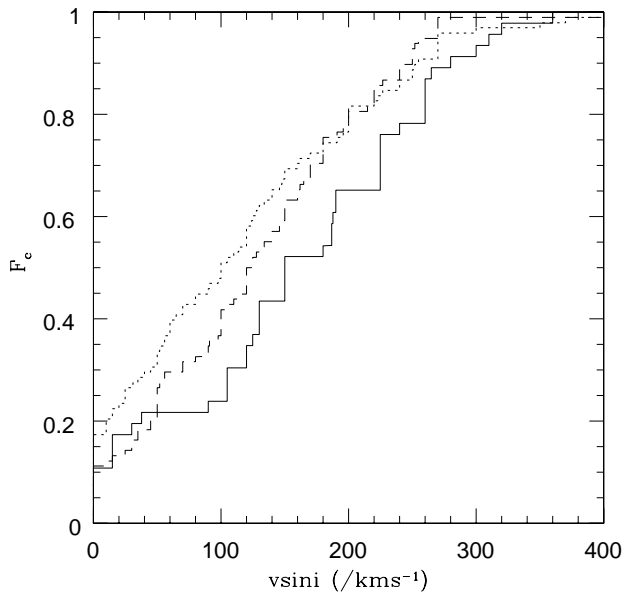,height=8cm}
\caption[The cumulative distribution of projected rotational velocities for
the Galactic cluster sample and for the LMC cluster sample.]{The distribution
of projected rotational velocities for the Galactic cluster sample (dotted
line) and for the LMC cluster sample (solid line). The dashed line is the
distribution for the Galactic cluster sample after simulation of the
systematic effects of low S/N (see text).}
\label{kstestgclmcc}
\end{figure}

Let us now compare the distributions of $v$sin$i$ in both the LMC and Galactic
samples to investigate the possible role of metallicity in determining
$v$sin$i$. Figure \ref{kstestgclmcc} shows the cumulative distribution
functions of the cluster and Galactic field samples (solid and dotted lines
respectively). As discussed in section 2.1, the accuracy of our determinations
of $v$sin$i$ are critically limited by the obtainable S/N. To
statistically account for the systematic effects imposed by low S/N I have
taken the spectral type, $v$sin$i$ data of the Galactic field sample and
produced a set of synthetic spectra with a distribution of S/N mimicking that
of the LMC sample and have then determined the $v$sin$i$ distribution of
the population as for the LMC sample. This is shown in figure
\ref{kstestgclmcc} as the dashed line. As expected, the net effect is to shift
a proportion of the slowest rotators to higher $v$sin$i$.

The resulting K-S test between the LMC cluster sample and the S/N degraded
Galactic cluster sample rates the probability of these two samples been alike
at 6\% (D$^{+}$=0.22, N$_{1}$=49, N$_{2}$=151). This is suggests that the
average rotation rate is higher amongst the LMC cluster stars. Table
\ref{table3} summarises our results.

\begin{table}
\caption[]{mean $v$sin$i$ for the four samples discussed above.}
\begin{center}
\begin{tabular}{cc}
\hline
Location& mean $v$sin$i$ (kms$^{-1}$)\\
\hline
Galactic field & 85\\
Galactic young clusters & 116\\
LMC field & 112\\
LMC young clusters & 146\\
\hline
       
\hline
\end{tabular}
\end{center}
\label{table3}
\end{table}

\section{Discussion --- the role of evolution and metallicity} 

Our observations present us with two important conclusions: 1. Young cluster
stars (B0-2 V-III within 2 mag.\ of the MS terminus) are faster rotators than
a similar luminosity range in the field, and 2. LMC cluster stars are faster
rotators than Galactic cluster stars within the same luminosity range at just
under a 2$\sigma$ significance.

To account, firstly, for the observed difference in rotational velocity
between the cluster and field populations I propose a scenario of
evolutionary rotational enhancement. This scenario utilizes the one clear
difference between the cluster and field populations, that is, the age spread
within each population.

Consider a spin-up phase which occurs over an interval of the MS
lifetime towards the end of the MS. The small age spread in the cluster
population at the luminosity of the cluster MS terminus places the majority of
these stars within the interval of spin-up. The field population, on the other
hand, possesses a more uniform spread of ages at a given luminosity, from the
ZAMS to the MS terminus (this can be confidently said for such rapidly
evolving stars). Hence in the field we see the time-average of the MS
rotational velocity.

The evolution of rotational velocity has been examined by Endal \& Sofia
(1979) and more recently, by Heger, Langer \& Woosley (2000) and Meynet \&
Maeder (2000). These models explicitly follow the radial exchange of angular
momentum during the course of stellar evolution. 

The studies of Heger, Langer \& Woosley (2000) and Meynet \& Maeder (2000)
focus on stars of $\ge$20M$_{\odot}$ which is more massive than that
considered in our sample (12-5M$_{\odot}$; Bressan et al. 1993). A major
distinction between massive stars and those considered in our sample is the
dominance of mass loss on the course of MS evolution. Their high mass-loss
rates lead to the removal of large amounts of angular momentum. Mass-loss is
enhanced amongst fast rotators that approach the critical angular velocity
. The removal of large amounts of mass from a $\ge$20M$_{\odot}$ star
results in a decline in the angular velocity over the course of MS life.

The study of Endal \& Sofia (1979) focuses on the rotational velocity
evolution of a 5M$_{\odot}$ star. Here, the mass-loss rates are much
reduced. The authors find that stars commencing their MS lives with relatively
slow angular velocities remain slow rotators throughout the course of the MS
evolution (although there is a marked increase in angular velocity of all
stars during the core contraction phase at the exhaustion of central hydrogen
burning this phase is far too short to account for any observable increase in
the number of rapid rotators). By contrast, stars commencing their lives with
an angular velocity greater than 60-80\% of the critical angular velocity spin
up towards the critical velocity over a moderate fraction of the MS lifetime.

The largest masses considered here are of the order of 12M$_{\odot}$. Such
stars have a mass loss rates 1.5 dex lower than those of a 20M$_{\odot}$ star
(de Jager et al.\ 1988) and consequently the MS evolution is not dominated by
mass-loss. Hence we can assume the behaviour of the present sample is best
described by the models of Endal \& Sofia.

In a cluster, therefore, those stars closer to the MS turnoff are more likely
to be rapid rotators since a proportion of them will have been spun up by
evolution. Less luminous stars that have not evolved as far through the MS
phase will not have entered the spin up phase. This process could provide the
mechanism required to explain the more rapidly rotating sample seen in the
cluster population in the vicinity of the MS turnoff.

Our second finding, that the LMC cluster and field samples exhibit more rapid
rotation than their Galactic counterparts indicates a metallicity effect. Our
sample is insufficient to determine whether the metallicity dependence resides
in the initial distribution of rotational velocities or if the magnitude of
the evolutionary spin-up discussed above increases in lower metallicity
environments. Our observational sample is necessarily limited by the small
sample size attainable with our instrumentation. To verify the findings
presented here we await the extension to a larger LMC (and SMC) sample. This
is currently underway utilising the FLAMES multi-object spectrograph on the
VLT.

\section{Summary}

In this paper I have presented a comparison of the rotational velocities of a
sample of stars within the field and from a selection of young clusters in
both the Galaxy and, from our own observations, the LMC. I have presented the
distribution of 100 LMC early B-type main-sequence stars. This represents the
first extragalactic study of the distribution of stellar rotational velocities.

It is found that the early B-star population of young clusters
(1-3$\times$10$^{7}$ yrs.) in both the Galaxy and the LMC exhibits more rapid
rotation than the field population. I propose this can be explained by a
scenario of evolutionary enhancement of the surface angular momentum brought
about by angular momentum redistribution over the main-sequence lifetime.

A comparison of a sample of field and cluster stars drawn from both LMC and Galactic environments shows that the LMC stars a more rapid rotators than their Galactic counterparts. The origin of this metallicity dependence is as yet unknown.

\section*{References}






\reference Abt, H.~A.~\and Hunter, J.~H.\ 1962, ApJ, 136, 381
\reference Bernacca, P. \and Perinotto, M. 1971, Contr.\ Oss.\ Astrofis.\
Asiago 239, 1
\reference Bressan, A., Fagotto, F., Bertelli, G., \and Chiosi, C.\ 1993, A\&AS, 100, 647 
\reference Brown, A.G.A. \and Verschueren, W. 1997, A\&A, 295, 63
\reference Burki, G. \and Maeder, A. 1977, A\&A, 57, 401
\reference de Jager, C., Nieuwenhuijzen, H. \and van der Hucht, K.A. 1988, A\&AS, 72, 259
\reference Endal, A.S. \and Sofia, S. 1979, ApJ, 232, 531
\reference Gies, D.R. \and Lambert, D.L.\ 1992, ApJ, 387, 673
\reference Gonz{\' a}lez Delgado, R.M. \and Leitherer, C. 1999, ApJS, 125, 479
\reference Gray, D. 1976, in The Observation and Analysis of Stellar Photospheres, (New York: Wiley)
\reference Guthrie, B.N.G. 1984, MNRAS, 210, 159
\reference Heger, A., Langer, M. \and Woosley, S.E. 2000, ApJ, 528, 368
\reference Hoffeit, D. \and Jaschele, C. 1982, The Bright Star Catalog, 4th ed. (New Haven: Yale University Observatory) 
\reference Hoffeit, D., Saladyga, M. \and Wlasuk, P. 1988, A Supplement to the Bright Star Catalog. (New Haven: Yale University Observatory)
\reference Keller, S.C., Wood, P.R. \and Bessell, M.S. 1999, A\&AS, 134, 489
\reference Keller, S.C., Bessell, M.S. \and Da Costa, G.S. 2000, AJ, 119, 1748
\reference Keller, S.C., Bessell, M.S. \and Da Costa, G.S. 2001a, AJ, 121, 905
\reference Keller, S.C., Grebel, E.K., Miller, G.J. \and Yoss, K.M. 2001b, AJ, 122, 248
\reference Keller, S.C. \and Wood, P.R. 2002, ApJ, 578, 144
\reference Korn, A.J., Becker, S.R., Gummersback, C.A. \and Wolf, B. 2000, A\&A, 353, 655
\reference Korn, A.J., Keller, S.C., Kaufer, A., Langer, N., Przybilla, N., Stahl, O., \and Wolf, B.\ 2002, A\&A, 385, 143
\reference Lennon, D.J., Dufton, P.L., Mazzali, P.A., Pasian, F. \and Marconi,
G. 1996, A\&A, 314, 243
\reference McErlean, N.D., Lennon, D.J., \and Dufton, P.L.\ 1999, A\&A, 349, 553
\reference Maeder, A., Grebel, E.K. \and Mermilliod, J.-C. 1999, A\&A, 346, 459
\reference Mermilliod, J.-C.\ 2000, in Stellar Clusters and Associations:
Convection, Rotation, and Dynamos, ASP Conf.~Ser.~198, (San Francisco: ASP), 514 
\reference Meynet, G. \and Maeder, A. 2000, A\&A, 361, 101
\reference Robertson, J.W. 1974, ApJ, 191, 67
\reference Rolleston, W.R.J., Brown, P.J.F., Dufton, P.L., \and Howarth, I.D.\ 1996, A\&A, 315, 95
\reference Royer, P. 2003 in Proceedings of the IAU Symposium 215 on Stellar Rotation, Cancun, Mexico, November 11-15 2002, ed. P. Eenens \& A. Maeder
\reference Slettebak, A. 1968, ApJ, 154, 933
\reference Slettebak, A., Collins G.W., Parkinson, T.D., Boyce, P.B. \and White, N.M. 1975, ApJS, 29, 137
\reference Townsend, R.H.D., Owocki, S.P.\ \and Howarth, I.D. 2004, submitted MNRAS, astro-ph:0312113
\reference Venn, K.A. 1995, ApJ, 449, 839
\reference Venn, K.A. 1999, ApJ, 518, 405
\reference von Zeipel, H. 1924, MNRAS, 84, 665
\reference Wolff, S.C., Edwards, S.\ \and Preston, G.W. 1982, ApJ, 252, 322
\reference Zorec, J. \and Briot, D. 1991, A\&A, 245, 150
\reference Zorec, J. \and Briot, D. 1997, A\&A, 318, 443



\end{document}